\documentclass[fleqn,usenatbib]{mnras}

\usepackage{graphicx}	
\usepackage{amsmath}	
\usepackage{amssymb}	
\usepackage{longtable}

\title[Swift J1357.2-0933 as seen with \emph{Swift} and \emph{NuSTAR}]{The black hole X-ray transient Swift~J1357.2-0933 as seen with \emph{Swift} and \emph{NuSTAR} during its 2017 outburst}

\author[Aru Beri  et.al. ]{Aru Beri$^{1,2}$ \thanks{a.beri@soton.ac.uk},
B.E.~Tetarenko$^{3,4}$,  
A. Bahramian$^{5,6}$,
Diego~Altamirano$^2$,
Poshak~Gandhi$^2$
\newauthor
G.R. Sivakoff$^3$,
N. Degenaar$^7$, M. J. Middleton$^2$, R. Wijnands$^7$,
J. V. Hern\'andz
\newauthor
Santisteban$^7$,~John A. Paice$^2$  \\
 $^1$DST-INSPIRE Faculty, Indian Institute of Science Education and Research (IISER) Mohali, 
Punjab 140306, India  \\
$^{2}$Physics \& Astronomy, University of Southampton, Southampton, Hampshire SO17 1BJ, UK\\
$^{3}$Department of Physics, University of Alberta, CCIS 4-181, Edmonton, AB T6G 2E1, Canada \\
$^{4}$Department of Astronomy, University of Michigan, 1085 South University Avenue, Ann Arbor, MI 48109, USA \\
$^{5}$Department of Physics and Astronomy, Michigan State University, East Lansing, MI, USA \\
$^{6}$International  Centre  for  Radio  Astronomy  Research,  Curtin
University, GPO Box U1987, Perth, WA 6845, Australia \\
$^{7}$Anton Pannekoek Institute for Astronomy, University of Amsterdam, Science Park 904, NL-1098 XH Amsterdam, the Netherlands \\
}

\begin{document}
%
%
%
\maketitle

\label{firstpage}
\begin{abstract}

We report on observations of black hole Swift J1357.2-0933, a member of the modest population of very faint X-ray transients. This source has previously shown intense dips in the optical lightcurve, a phenomena that has been linked to the existence of a ``unique toroidal structure'' in the inner region of the disc, seen at a high inclination. Our observations, carried out by the Neil Gehrels Swift and NuSTAR X-ray observatories, do not show
the presence of intense dips in the optical light curves.
~We find that the X-ray light curves do not show any features that would straightforwardly support an edge-on configuration or high inclination configuration of the orbit.~This is similar to what was seen in the X-ray observations of the source during its 2011 outburst. Moreover, the broadband spectra were well described with an absorbed power-law model without any signatures of cut-off at energies above 10~keV, or any reflection from the disc or the putative torus. Thus, the X-ray data do not support the unique ’obscuring torus’ scenario proposed for J1357. We also performed a multi-wavelength study using the data of X-ray telescope and Ultraviolet/Optical Telescope aboard \emph{Swift}, taken during the ∼4.5 months duration of the 2017 outburst.~This is consistent with what was previously inferred for this source.~We found a correlation between the simultaneous X-ray and ultraviolet/optical data and our study suggests that most of the reprocessed flux must be coming out in the ultraviolet.

\end{abstract}


%
%
%
%

\section{Introduction}\label{sec:intro}
Low mass X-ray Binaries~(LMXBs) are one of the brightest sources of 
X-ray emission in the sky and harbor a black hole~(BH) or neutron star~(NS).~These compact objects accrete gas from their low-mass companion star, typically having mass less than a solar mass~($M_{\odot}$).~However, in some BH-LMXBs the companion has a mass which is a few times $M_{\odot}$~\citep[e.g., GRS~1915+105;][]{Reid14}.
Most LMXBs are transients spending most of their time in quiescence with X-ray luminosities of $L_{X}{\sim}10^{30}-10^{33} \mathrm{ergs~s^{-1}}$,~interrupted by occasional outburst episodes. During outbursts, LMXBs 
accrete at much higher rates~(than when in quiescence), resulting in an increase in the observed $L_{X}$ by several orders of magnitude.
Transient LMXBs can be classified as bright X-ray transients, Faint X-ray transients, and Very faint X-ray transients~(VFXTs),
based on their 2-10~keV peak luminosity~($L_{X}^{peak}$) reached during outburst.
VFXTs are defined as the systems that reach $L_{X}^{peak}$ of only
$\sim$ $10^{34}-10^{36} \mathrm{ergs~s^{-1}}$, orders of magnitude
lower than bright X-ray transients, which display $L_{X}^{peak}{\sim}10^{37}-10^{39} \mathrm{ergs~s^{-1}}$ \citep[see][for details]{Wijnands06}.~This classification in luminosity classes is  phenomenological and does not necessarily reflect any physical difference between the systems. \\

Despite the growth of the known population of VFXTs over the last 15 years, the X-ray behavior of such systems is not yet completely understood. An example of a BH VFXTs is the Galactic LMXB, Swift~J1357.2-0933 (hereafter, J1357). 
This source was discovered on 2011, January 28 when an outburst was detected with  the \emph{Swift}~Burst Alert Telescope \citep{Krimm11a}. 
This outburst lasted for about 7~months. 
Several X-ray as well as ground-based
observations were carried out during this outburst \citep[see e.g.,][]{Casares11, Krimm11b, Milisavljevic11, Armas13a, Armas13b}. The peak X-ray luminosity~($L_{X}^{peak}$)~observed in 2-10~keV band during this outburst was ${\sim}10^{35} \mathrm{ergs~s^{-1}}$, placing 
J1357 in the category of VFXTs \citep{Armas13a}.
Optical observations performed during the 2011 outburst
revealed the black hole nature of the accretor \citep{Corral13}, making J1357 one of the very few VFXTs confirmed to harbor a BH  \citep{Armas13b,Corral13}.~Based on optical data,~\citet{Corral13} were able to find an orbital period of
2.8~h,~and based on optical observations the source distance is estimated to lie within a range of 0.5 and 6.3~kpc \citep{Rau11,Shahbaz13}. \\

The optical light curves also showed the presence of intense dips \citep{Corral13}. The optical dip recurrence time evolved over time and the frequency associated with the dip recurrence time was found to decrease throughout the outburst decay \citep{Corral13}.
The dips observed in the optical light curves have been linked to
the existence of a ``toroidal structure'' in the  inner region of the disc.~This obscuring structure is thought to move outwards as the mass accretion rate decreases and can only be seen at very high inclination \citep[$i$$>$70$^{\circ}$,][]{Corral13,Mata15}.
The light curves of J1357 during 
quiescence showed similar dips and also flares \citep{Shahbaz13,Russell17}.
This kind of large optical variability has never been observed in any other LMXB when they are observed at similar time resolution \citep{Russell17}. \\

LMXBs that have $i$ between 60$^{\circ}$ and 80$^{\circ}$ with respect to the line of sight are observed to show signatures of the orbital motion~(eclipse) or dips in the X-ray light curves \citep[see e.g.,][]{Parmar86,Courvoisier86}.~The X-ray observations made with \emph{Rossi X-ray Timing Explorer}~(\emph{RXTE}) and \emph{XMM-Newton} during the outburst of J1357 in 2011 did not show any of these features \citep{Corral13,Armas13b}. 
The lack of signatures of the orbital motion in the \emph{RXTE} light curves were suggested to be because of the existence of extreme mass ratio in this system \citep{Corral13}.~The very small size of the donor could make the X-ray eclipse very shallow and difficult to detect in J1357.~However, this  interpretation
is not secure and not all the observed phenomena can be explained by it \citep[see][for a discussion]{Armas13b}. \\

During the 2011 outburst of J1357, one of the \emph{RXTE} observations made near the beginning of the outburst revealed the presence of a quasi-periodic oscillation~(QPO) at a frequency~(milli-hertz; mHz) similar to  that of the optical dips. However,~this feature was not present in any of the subsequent X-ray observations made close to the dates of the detection of the optical dips \citep{Armas13b}.
The same authors found that on the extrapolation of the frequency of optical dips near the beginning of the outburst resulted into a much higher frequency than observed in X-rays.
Thus, it is not clear whether the mechanism behind the optical dips is similar to that behind the X-ray QPO \citep{Armas13b}. \\

The X-ray spectrum of J1357 has also been investigated during its outburst in 2011
with data from the \emph{Swift} and \emph{XMM-Newton} observatories
\citep{Armas13a, Armas13b}.
The \emph{Swift}-\textsc{XRT} observations showed 
softening of the X-ray spectra with the decrease in X-ray flux. This behavior is typical of X-ray binaries that accrete at sub-luminous~accretion rates \citep[$\sim$ $10^{34}-10^{36} \mathrm{ergs~s^{-1}}$; see][]{Armas13a,Wijnands15}.
The high quality X-ray data obtained with \emph{XMM-Newton} showed the presence of soft thermal disc component with a temperature of $\sim$0.22~keV and a hard Comptonized component~(\textsc{NTHCOMP}) with a photon index of $\Gamma$ $\sim$ 1.6 and an electron temperature~($kT_e$) $\sim$ 8.2~$\mathrm{keV}$
\citep{Armas13a}.
However, no evidence of reflection features that support the torus scenario were found in the X-ray spectrum. Thus, the overall geometry of J1357 is still under debate. \\

On 2017 April 20,~J1357 underwent its second outburst \citep{Drake17,Sivakoff17} providing an opportunity to further investigate this source.
One of the ways to test the putative torus model is to study the broadband X-ray spectrum covering energies above 10~keV. The detection of any signatures of a high energy cut-off,
or any reflection from the disc will be quite useful to test the torus model. Here, we report results obtained using data from the \emph{Swift} and \emph{NuSTAR} observatories. \\

\begin{table*}
\caption{Log of the \emph{Swift} observations and \textsc{XRT} spectral results}
\begin{tabular}{ c c c c c c c c c}
\hline
\hline
Obs-ID & Time~(MJD) & Mode  & Exp-time~(ksec) & $\Gamma$ & $F_{\rm{X,unabs}}^{a}$ & $L_{\rm{X}}^{b}$ &  ${\chi}^2_{\nu}$~(dof)   \\
\hline
\hline
31918049 & 57864.8471  & WT & 1.4 & $1.45\pm0.14$   & $15.0\pm{1.0}$   & $4.0\pm0.2$   & 0.53~(27) \\
31918050 & 57866.7112 & WT   & 1.4 & $1.46\pm0.09$   & $14.4\pm{0.8}$  &  $3.8\pm0.2$  & 0.71~(20) \\
31918051 & 57867.6940 & WT    & 0.7 & $1.47\pm0.07$   & $15.0\pm{0.6}$   &    $4.0\pm0.1$ & 0.70~(37)\\
88094002 & 57871.1618 & PC    & 1.0 & $1.49\pm0.08$   & $12.7\pm{0.6}$   &    $3.4\pm0.1$ & 0.94~(46)\\
31918054 & 57874.8192 & WT &  0.2 & $1.60\pm0.13$ & $14.0_{-1.5}^{+1.6}$ & $3.8\pm0.4$ & 0.73~(24) \\ 
31918055 & 57876.2725  & WT & 0.5 & $1.64\pm0.09$  & $11.8\pm0.9$ & $3.2\pm0.2$  & 1.2~(24) \\
31918056 & 57878.3949 & WT & 0.5 & $1.54\pm0.08$ & $13.3\pm0.9$  & $3.5\pm0.2$  & 1.2~(32) \\
31918057 & 57885.7127 & WT  & 1.0 & $1.57\pm0.04$ & $8.9\pm0.3$  & $2.4\pm0.2$  & 1.06~(76) \\
31918058 & 57888.5048   & WT  & 1.0  & $1.61\pm0.09$ & $8.0\pm0.6$  & $2.2\pm0.2$  & 1.07~(49)\\
31918059 & 57893.0720 & WT & 0.6 & $1.69\pm0.10$ & $6.4\pm0.5$  & $1.7\pm0.1$  & 1.0~(39)\\
31918060 & 57894.8105  & WT  & 1.0 & $1.58\pm0.08$ & $7.5\pm0.5$  & $2.0\pm0.1$   & 0.44~(23) \\
31918061 & 57896.3281 & WT  & 1.0  & $1.68\pm0.07$ & $7.3\pm0.4$ & $1.9\pm0.1$  & 0.97~(34)\\
31918062 & 57901.3109 & WT & 1.3  & $1.62\pm0.07$ & $6.2\pm0.3$ & $1.7\pm0.1$ & 1.01~(79) \\
31918064 & 57904.3665 & WT  & 0.8  & $1.57\pm0.10$ & $5.5\pm0.4$ & $1.5\pm0.1$  & 0.84~(37) \\
31918065 & 57906.1601 & WT & 0.80  & $1.57\pm0.10$ & $5.8\pm0.5$ & $1.5\pm0.1$  & 1.1~(40)\\
31918066 & 57914.6150 & WT  & 0.7  & $1.68\pm0.12$ & $4.3\pm0.4$ & $1.1\pm0.1$   & 0.91~(27)\\
88197001 & 57915.1940 & PC  & 1.6 & $1.53\pm0.08$ & $3.9\pm0.3$ & $1.10\pm0.08$  & 0.67~(36)\\
31918068 & 57921.5100  & WT & 1.0  & $1.65\pm0.12$ & $3.2\pm0.3$ & $0.84\pm0.08$  & 0.86~(29) \\
31918069 & 57924.6414 & WT & 1.1    & $1.65\pm0.09$ & $3.4\pm0.3$ & $0.92\pm0.07$  & 0.77~(36) \\
31918070 & 57925.8951 & WT &  1.7  & $1.63\pm0.07$ & $2.9\pm0.1$ & $0.81\pm0.03$  & 0.99~(27) \\

31918071 & 57932.1508 & WT  & 1.8   & $ 1.75\pm0.07$ & $2.43\pm0.08$    &  $0.72\pm0.02$ & 1.35~(49) \\

31918072 & 57934.2056 & WT &  1.0   & $1.71\pm0.15$  & $2.0\pm0.3$   & $0.54\pm0.06$  & 0.53~(18)\\

31918073 & 57938.9768  & WT & 1.4   & $1.71\pm0.08$  & $1.95\pm0.09$   &  $0.52\pm0.02$ & 1.2~(26) \\

31918074 & 57944.0258 &  WT  & 1.4   & $1.76\pm0.08$ & $2.08\pm0.08$   & $0.55\pm0.02$  & 0.94~(33) \\

31918075 & 57949.2266  & WT &  1.3  & $1.77\pm0.09$  & $1.7\pm$0.1  &	$0.45\pm0.03$  & 0.80~(24)\\

31918076 & 57954.5955 & WT & 2.0  & $1.8\pm0.1$  & $1.20\pm0.05$       & $0.32\pm0.01$   & 1.21~(29) \\

31918077 & 57959.2528 & WT & 1.8 & $1.87\pm0.10$ & $1.09\pm0.05$& $0.29\pm0.01$   & 0.73~(25)\\

31918078 & 57969.7412 & PC & 2. & $1.74\pm0.13$ & $0.5\pm0.1$ & $0.15\pm0.03$   & 1.13~(21) \\
31918079 & 57974.4516 & PC & 2.0 & $1.70\pm0.14$  & $0.8\pm0.1$ & $0.22\pm0.03$   & 1.08~(15)\\
31918080 & 57979.1890 & PC & 1.0 & $ 1.90\pm0.28$ & $0.5\pm0.1$ & $0.14\pm0.03$   & 1.31~(7)\\
31918082 & 57997.1966 & PC  & 0.7 & $1.75\pm0.33$  & $0.4\pm0.1$ & $0.10\pm0.03$   & 1.30~(3)\\

\hline
\hline
\end{tabular}
\label{Swift}
\\
{Note. $^a$~Flux~($F_{\rm X,unabs}^{\rm a}$) is in units of $10^{-11} \rm ergs~\rm cm^{-2}~\rm s^{-1}$. 
$^b$~X-ray luminosity~($\rm L_{\rm X}^{\rm b}$) in units of $10^{34}~\rm ergs~\rm s^{-1}$~calculated from the 0.5--10 keV unabsorbed flux by adopting a distance of 1.5~kpc.}
\end{table*}

\begin{figure}
\includegraphics[height=0.75\columnwidth]{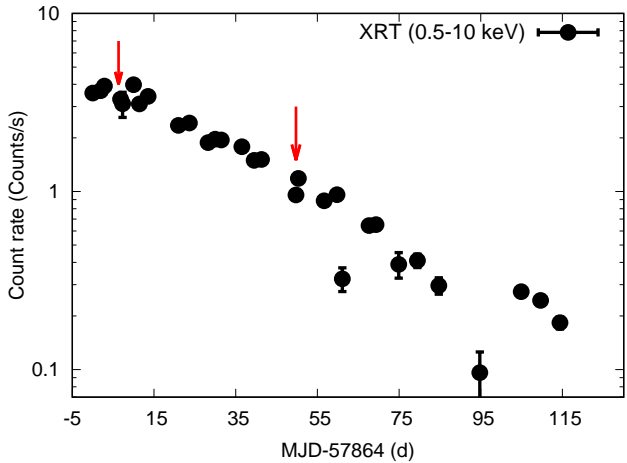}
\caption{This plot shows the count rate curve obtained using the \emph{Swift}-\textsc{XRT} during the 2017 outburst
of Swift~J1357.2-0933 which began on $\sim$57864~MJD. The arrows marked in red indicate the dates~(mentioned in the text) during which the \emph{NuSTAR} observations were made.~The y-axis is plotted in log scale.}
\label{outburst}
\end{figure}

\begin{figure*}
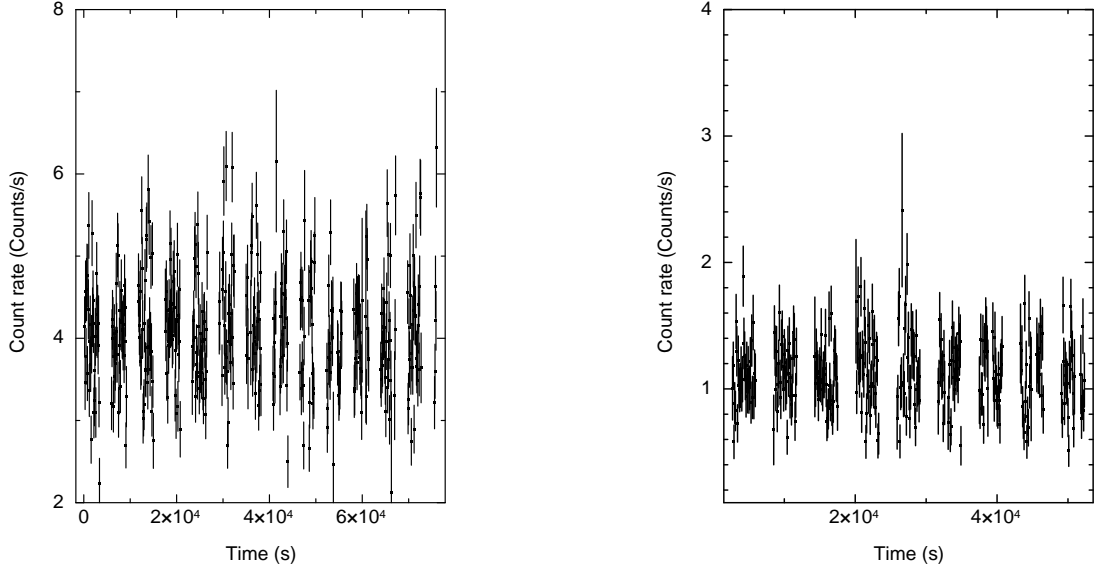

\centering
\begin{minipage}{0.45\textwidth}
\includegraphics[height=\columnwidth,angle=0]{fig2.ps}
\end{minipage}
\hspace{0.02\linewidth}
\begin{minipage}{0.45\textwidth}
\includegraphics[height=\columnwidth,angle=0]{fig3.ps}
\end{minipage}
\caption{Left: Light curve obtained using the data of FPMA of \emph{NuSTAR} observation made on 2017 April 28.~Right:~Light curve created using the data of FPMA during the second \emph{NuTSAR} observation made on 2017 June 10.~The plotted light curves are binned with a bin size of 100~s. }
\label{LC}
\end{figure*}

\begin{figure*}
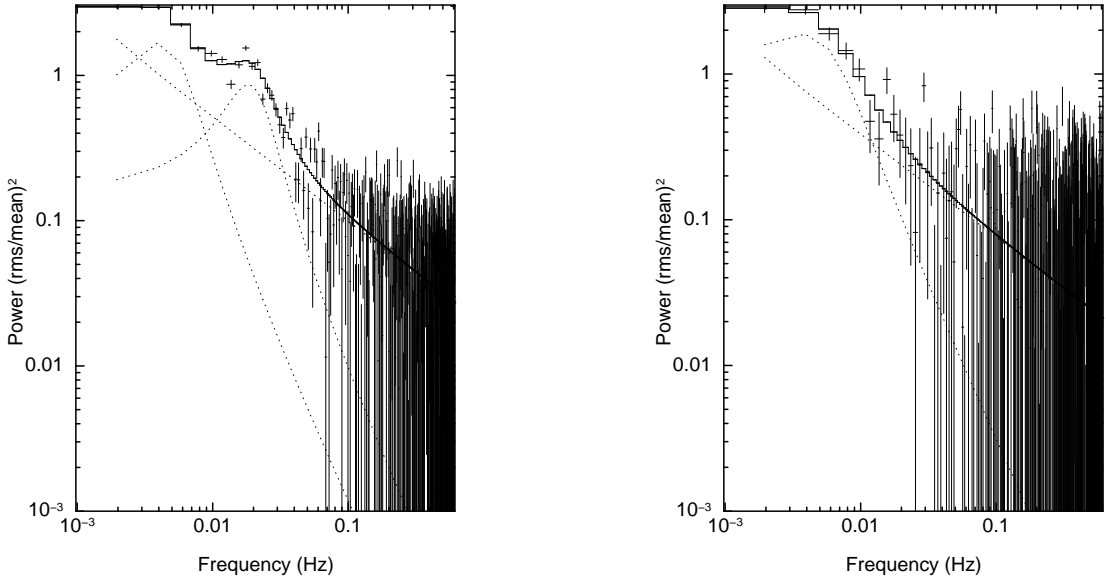

\centering
\begin{minipage}{0.45\textwidth}
 \includegraphics[height=\columnwidth,angle=0]{fig4.ps}
\end{minipage}
\hspace{0.02\linewidth}
\begin{minipage}{0.45\textwidth}
\includegraphics[height=\columnwidth]{fig5.ps}
\end{minipage}
\caption{Left:~The Cross Power Density Spectrum obtained using the data of FPMA and FPMB of \emph{NuSTAR} observation made on 2017 April 28.~Right:~The Cross Power Density Spectrum created using data of both focal modules~(A \& B) of observation made on 2017 June 10.}
\label{powspec-1}
\end{figure*}

\begin{figure}
\includegraphics[height=4.5in,width=4.5in,angle=0,keepaspectratio]{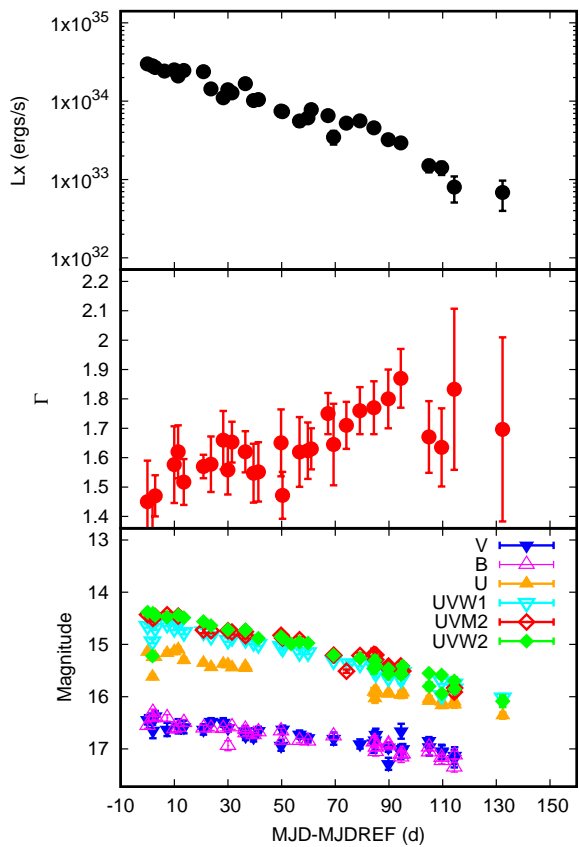}
\caption{The upper panel is the light curve in the 0.5--10~keV band obtained using the \textsc{XRT} observations (assumed source distance of 1.5 kpc), the middle panel is the photon index~($\Gamma$) evolution with time while in the bottom
panel the UV/optical light curves in the Vega system are plotted.~The reference time~(MJDREF) used for the 2017 outburst is,~57864.84~MJD.}
\label{PL-index}
\end{figure}

\begin{figure*}
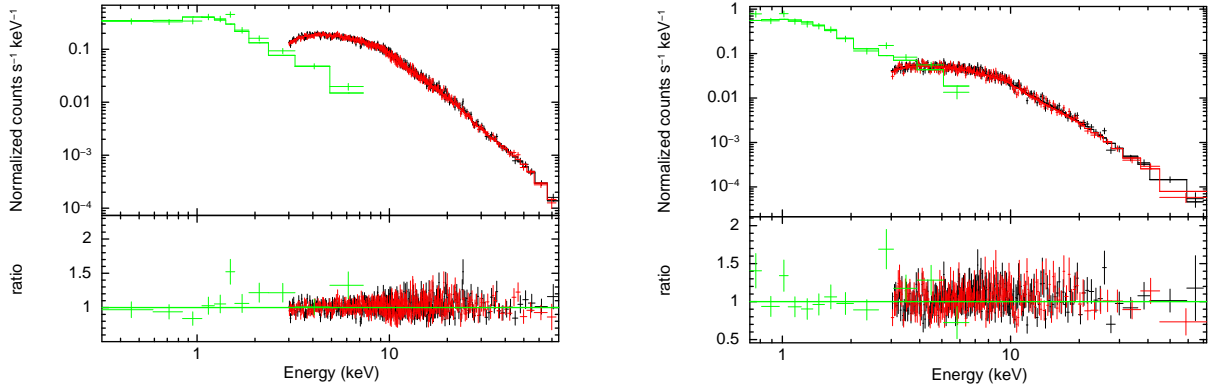

\centering
\begin{minipage}{0.45\textwidth}
\includegraphics[height=3.in,width=7.5cm,angle=-90,keepaspectratio]{fig7.eps}
\end{minipage}
\hspace{0.02\linewidth}
\begin{minipage}{0.45\textwidth}
\includegraphics[height=3.in,width=7.5cm,angle=-90,keepaspectratio]{fig8.eps}
\end{minipage}
\caption{(left)~We show the best-fit spectrum obtained using quasi-simultaneous data from \emph{Swift} in PC mode (Obs~ID--00088094002) and \emph{NuSTAR}~(Obs~ID--90201057002).~(right)~Best-fit spectrum obtained using data from \emph{Swift} in WT mode (Obs~ID--00031918066) and the second \emph{NuSTAR} observation~(Obs~ID--90301005002) }
\label{spec}
\end{figure*}

\section{Observations and data analysis}\label{sec:dataanalysis}
\subsection{NuSTAR}
The Nuclear Spectroscopic Telescope Array (\emph{NuSTAR}) mission is the first X-ray satellite with multi-layer hard X-ray optics. It operates in the 3 to 79~keV energy band \citep{Fiona13}. 
There are two identical telescopes aboard with grazing incidence optics, each one focusing on separate detector modules, Focal Plane Modules~A and B (\textsc{FPMA} \& \textsc{FPMB}), at a distance of 10~m. 
In addition to its imaging capabilities that extend well above 10~keV, \emph{NuSTAR} has a time resolution of 10~${\mu}$s, which allows us to study variability in LMXBs.

During the 2017 outburst of J1357, two \emph{NuSTAR} observations were performed. One of the observations was made  between 12:51 \textsc{UT} on 2017~April  28 and 10:06  \textsc{UT} on 2017~April  29,~while the second
observation was performed between 13:36 \textsc{UT} 2017~June 10 and  03:46 \textsc{UT} on 2017~June 11.~We have used both observations in this work.~The standard processing and the extraction were performed using  \textsc{HEASOFT} v6.19 and \textsc{NUSTARDAS}~(V1.9.1). The light curves, spectra, and the response files were
extracted using \textsc{NUPRODUCTS}.
We have used a circular region of 30 arc-second radius for the source and a void circular region of 30 arc-second radius on the same chip for the background files.~J1357 was detected well above the background
in the entire \emph{NuSTAR} energy band.

\subsection{Swift}
The  Neil Gehrels \emph{Swift} observatory~\citep{Gehrels04}, launched in November 2004, has three instruments on board: a) the Burst Alert Telescope~(\textsc{BAT}) operating in the energy range of 15-150~keV \citep{Barthelmy05}, b) the X-ray Telescope~(\textsc{XRT}) which works in soft X-ray band~\citep[0.2-10~keV;][]{Burrows05} and c) the Ultraviolet and Optical Telescope~(\textsc{UVOT}), which covers UV and optical bands  \citep[70-600~nm;][]{Roming04}. \\

We monitored J1357 for $\sim4.5$ months (30 pointings totaling $\sim35$ ks) over the course of its 2017 outburst with \emph{Swift}, using the XRT and the UVOT.
A log of \emph{Swift}-\textsc{XRT} observations is given in Table~\ref{Swift}.
Two of the \emph{Swift} observations are contemporaneous with the \emph{NuSTAR} observations.
One of them was made between 14:49 and 16:38 \textsc{UT}
on 2017 April 28~(Obs ID~00088094002) and the other was performed between
14:44 and 14:55 \textsc{UT} on 2017 June 10~(Obs ID~00031918066).
The \textsc{XRT} observation with ID~00088094002 was made in photon counting (PC) mode, while during the  other observation with ID~00031918066  data were collected in the windowed timing (WT) mode.
The average countrate in the 0.5--10~keV band during these observations is $3.3 {\pm} 0.1$ and $1.45\pm0.04$ $\mathrm{count~s^{-1}}$ respectively. \\

We reprocessed the \textsc{XRT} data using \textsc{xrtpipeline} and manually extracted source and background spectra for each observation using \textsc{xselect}~(from the Heasoft software package;~version~6.19).
Due to high source count rate, most of our observations were performed in WT, while a few were taken in PC.~For WT mode observations, we extracted spectra from only grade 0 events to avoid possible spectral residuals in the spectrum\footnote{e.g., see: http://www.swift.ac.uk/analysis/xrt/}.
The source events were obtained from a circular region using one of the two values of radius $\sim$~35~${\arcsec}$ or ~47~${\arcsec}$, depending on the source brightness. The background
events were extracted from the outer source-free regions, with a circular 
region of same radius as that used for the extraction of source events.
We investigated the PC mode observations for evidence of pile up following procedures provided in the Swift/XRT guide\footnote{http://www.swift.ac.uk/analysis/xrt/pileup.php}. 
We found pile-up issues in the two observations~(ID~00088197001 and 00031918072).~For these two observations, we extracted the source spectrum from an annulus excluding the piled up region.
For the observation with ID~00088197001, the annulus has inner and outer radii of 7${\arcsec}$ and 47${\arcsec}$, and for the other observation (ID~00031918072) they were 5${\arcsec}$ and 35${\arcsec}$, respectively.~For the PC mode data which did not exhibit any pile-up issues we have used a circular region of size $\sim$~35${\arcsec}$. For all the PC mode observations, the background was typically extracted from source-free $\sim$~200${\arcsec}$ radius circles.
We generated exposure maps for observations using the task \texttt{xrtexpomap} and created Ancillary Response Files (ARFs) using \texttt{xrtmkarf}.~Finally, we performed spectral analysis in the 0.3--10 keV band for PC mode data and the 0.7--10 keV band for WT mode data.  \\

All the \textsc{UVOT} observations were taken in image mode. The majority of the observations were taken in all six available
filters (v, b, u, uvw1, uvm2, uvw2), with a few exceptions. We used the \textsc{uvotmaghist} tool to create light curves, in each filter. This tool uses \textsc{uvotsource} to perform aperture photometry on all sky images (in each filter) available for an individual observation and calculate the source magnitude (in the Vega system) and flux densities. Aperture photometry was performed with a circular region with a radius of 5 ${\arcsec}$ centered on the source position. A neighboring source-free circular sky region with a radius of 10 ${\arcsec}$ was used for background correction.
We have corrected the magnitudes and fluxes for the Galactic extinction. 
The reddening $E(B-V)$ is 0.04 mag in the direction of J1357.~We have used the following values of extinction: $A_v$~=~0.123, $A_b$~=~0.163,~$A_u$~=~0.193,~$A_{uvw1}$~=~0.263,~$A_{uvm2}$~=~0.387,
$A_{uvw2}$~=~0.349,~which were obtained by \citet{Armas13a}.

\subsection{X-ray Spectral Analysis}

We performed X-ray spectral analysis using \textsc{xspec} 12.9.1 \citep{Arnaud96} and applied $\chi^2$ statistics.
The spectra obtained from both the detectors~(\textsc{FPMA} and \textsc{FPMB}) of \emph{NuSTAR} were grouped into bins with a minimum of 20 photons using \textsc{grppha}. For the case of spectra obtained with {\bf the} \textsc{XRT} we have grouped data with a minimum of 15 counts per bin.
Interstellar absorption was included in all our spectral fits, employing \citet{Wilms00} abundances of elements and \citet{Verner96} photo-electric cross-sections.
For all the observations performed with {\bf the} \textsc{XRT}, we used \textsc{phabs}
model and fixed $N_H$ to $1.2 \times 10^{20}~\rm{atoms~cm}^{-2}$ for the spectral fitting. This value 
was obtained by \citet{Armas13b} from the high-resolution X-ray
spectra obtained with \emph{XMM-Newton}.~By using the same absorption model,~abundances,~cross-sections as \citet{Armas13a} in our spectral analysis, we can directly compare the two different outbursts.
\emph{NuSTAR} works in the energy range of 3-79~keV, therefore, to study the low energy emission below 3~keV we have used the data from the \emph{Swift}-\textsc{XRT} observations made contemporaneous to the \emph{NuSTAR} observations.~For the case of simultaneous spectral fitting of \emph{Swift}-\textsc{XRT} and \emph{NuSTAR} observations we have used the latest and more updated model \textsc{tbabs} for taking into account the interstellar absorption.
We allowed the cross-calibration constant~(CC) between the instruments to take a difference in absolute flux calibration into account.
All the fluxes are given with respect to~\textsc{FPMA}.
 The values of the CC obtained for \textsc{FPMB} and \emph{Swift}--\textsc{XRT} during the first contemporaneous \emph{NuSTAR}~(ID~90201057002) and \emph{Swift}~(ID~00088094002)
 observation are $1.035\pm0.006$ and $0.75\pm0.03$ respectively.
 The observed values of the CC for the second contemporaneous 
\emph{NuSTAR}~(ID~90301005002) and \emph{Swift}~(ID~00031918066) observation are as follows:~$1.01\pm0.01$~(\textsc{FPMB})
 and $0.80\pm0.03$~(\textsc{XRT}).
Unless explicitly mentioned, we quote all errors at 1-$\sigma$ confidence level. \\


%
%
%
%

\section{Results}
\subsection{Timing Results}

In Figure~\ref{outburst}, we show the long-term 0.5-10~keV light curve of J1357 obtained from the \emph{Swift}-\textsc{XRT} observations. The \emph{NuSTAR} observations used in this work are marked in red.
Here, we note that \citet{Bailey18} 
found that during both outbursts observed for J1357 the  exponential~(viscous) decay timescale is about $\sim$~64~days. \\

To search for signatures of the orbital motion~(2.8~h) or the presence of dip-like features similar to that observed in the optical light curves \citep[e.g.,][]{Corral13}, we created light curves in 3-79~keV~band from the two \emph{NuSTAR} observations. 
In Figure~\ref{LC} we show the light curves obtained with the \textsc{FPMA} for each observation.
The average count rate measured during the first and the second \emph{NuSTAR} observation is $\sim$~4~$\mathrm{count~s^{-1}}$ and $\sim$~1~$\mathrm{count~s^{-1}}$ respectively.~We do not observe any dip-like features or the presence of eclipses in the X-ray light curves. \\

Power density spectral analysis was performed using the timing analysis software \textsc{HENDRICS 3.0}\footnote{https://github.com/StingraySoftware/HENDRICS} (\textit{High ENergy Data Reduction Interface from the Command Shell}).~This software is well suited to create the cross power density spectrum~(CPDS) which is a proxy for the power density spectrum~(PDS) that uses the signals from two independent detectors of \emph{NuSTAR} instead of a single one \citep[for details see][]{Bachetti15a}.~This software also allowed us to perform poisson noise subtraction.~The light curves in 3-79~keV band, having a bin size of $\sim$1~s were used to search for signals in the low frequency range.  \\

The CPDS obtained with the first \emph{NuSTAR} observation ~(ID-90201057002) was fitted using two zero-centered Lorentzian and a power law \citep{Belloni02}.
The obtained values of characteristic frequencies~($\nu_o$) of the two Lorentzian components are 
$4.1\pm0.2$ ~mHz and $18.3\pm0.4$~mHz and the corresponding full width at half maximum~(FWHM) values
are $4.9\pm0.8$~mHz and $19.9\pm0.2$~mHz, respectively.~Thus, the Q factor of the two Lorentzians at $\sim$ 4~mHz and 18~mHz  is $\sim$ 0.8 and 1, respectively (also given in Table~\ref{QPO}).
The root mean square~(rms) variability of the two Lorentzian components are  $0.055\pm0.007$ and $0.008\pm0.001$ $\%$, respectively (Figure~\ref{powspec-1}).
The value of the power-law index observed for the CPDS is $0.71\pm0.04$.~The second \emph{NuSTAR} observation~(ID-90301005002) could be well fitted with a Lorentzian and a power law. The observed values of the characteristic frequency and  the root mean square~(rms) variability of the Lorentzian component in the CPDS of the second \emph{NuSTAR} observation is $0.004\pm0.001$ and, $0.02\pm0.01$ $\%$, respectively. 
The FWHM of the feature at $\sim$~4~mHz is $\approx$~$0.007\pm0.002$ (and thus a Q factor of $\sim$ 0.6).~The value of the power law index is $0.7\pm0.1$.~In this second observation, we did not require any low frequency component around 0.0183~Hz, plausibly due to the lower count-rate in this observation. \\

In Table~\ref{QPO}, we summarize the values of the
frequencies observed with the two \emph{NuSTAR} observations during the
outburst of J1357 in 2017, along with the previous known value of QPO frequency during its outburst in 2011.~Here, we notice that during the 2017 outburst, J1357 exhibits low frequency variability, having fractional rms value quite low compared to its previous outburst and also to other black hole binaries in their hard state \citep[see][for details]{Belloni14}. \\

 \begin{table*}
\caption{Observed values of Frequencies~($\nu$) in the PDS of Swift~J1357.2-0933}
\begin{tabular}{cc |ccccccc}
 \hline
       &              &          &          \\ [0.5ex]       
  Outburst & Observatory & MJD &  $\nu$1~(mHz) & Q~Value & rms~($\%$) & $\nu$2~(mHz) & Q~Value  & rms~($\%$) \\ [0.5ex] 
\hline

2011& \emph{RXTE} & 55594 & $5.9\pm0.1$ & 3 & $12\pm3$ & none & - & -  \\ [0.5ex] 
2017 & \emph{NuSTAR} & 57871.54 & $4.1\pm0.2$ & $0.8\pm0.1$ & $0.008\pm0.001$ & $18.3\pm0.4$ & $0.915\pm0.001$ & $0.055\pm0.007$  \\ [0.5ex] 
2017 & \emph{NuSTAR} & 57914.57 & $4\pm1$ & $0.57\pm0.21$ & $0.02\pm0.01$ & none & - & -  \\ [0.5ex]
\hline

\end{tabular}
\\
{{\bf{Notes}}:  
Errors quoted are for the 90 $\%$ confidence range.}  \\
\label{QPO}  
\end{table*}

 \begin{table*}
\caption{Best Fitting Parameters Obtained Using an Absorbed Power law }
      \label{best-fit}
 \begin{tabular}{cc |cccc}
 \hline
       &              &          &       &     \\ [0.5ex]       
  \emph{NuSTAR} Obs~ID & N$_H$ (fixed) & $\Gamma$ & $N_{PL}^a$ & Flux  & Reduced ${\chi}^2$~(dof) \\ [0.5ex] 
\hline

90201057002 & $0.012$ & $1.663\pm0.005$ & $0.0235\pm0.0003$ & $3.50\pm0.01$ & 0.96~(1179)\\ [0.5ex] 
 90301005002 & $0.012$ & $1.79\pm0.01$ & $0.0084\pm0.0002$ & $0.89\pm0.04$ & 1.01~(595)\\ [0.5ex] 
\hline

\end{tabular}
\\  
{{\bf{Notes}}:  
  N$_{\rm{H}}$ is in units of 10$^{22}~\rm{atoms~cm}^{-2}$. 
 Unabsorbed flux in 0.3-79~keV band is in units $10^{-10} \rm{ergs~cm}^{-2}~\rm{sec}^{-1}$.~a $\rightarrow$ Power law normalization~($N_{PL}$) is in units of $\rm{photons~cm}^{-2}~\rm{s}^{-1}~\rm{keV}^{-1}$ at 1~keV.} \\

\label{Best-fit1}  
\end{table*}   

\begin{table*}
\caption{Spectral fit parameters with other phenomenological models. For each spectral fit we have used $N\rm{_{H}}$ = $0.012 {\times}10^{22} \rm{cm^{-2}}$.  For the first observation, the spectral fitting was performed in 0.3-79~keV band while for the second we have used 0.7-79~keV band.}
\label{compt}
\begin{tabular}{ c c c c c c c c c}
\hline
\hline
Parameters & \textsc{cutoffpl} & \textsc{bknpower} &  \textsc{diskbb+nthcomp}
& \textsc{diskbb+comptt}  \\
\hline
&   &  Observation~1 &  \\
\hline
$\Gamma$ &$1.63\pm0.01$   &   $1.57\pm0.05$  & $1.695\pm0.005$  & -\\
${\Gamma}_{2}$ & -  &   $ 1.685\pm0.007$     & - &      & -      \\
$E_{cut/break/e}$ (keV)  &  $301_{-74}^{+142} $        &  $5.4_{-0.9}^{+2.1}$  &  $46_{-12}^{+67}$  & $32_{-7}^{+11}$\\
$kT\rm_{in}~(keV)$ & - & - & $0.3\pm0.3$ & $0.10\pm0.03$ \\ 
$\tau$  & -  &  - & - & $1.6\pm0.3$  \\
$N^{a}$ & $0.0224\pm0.0004$ & $0.020\pm0.001$ & $0.0239\pm0.0002$ & $0.0015\pm0.0003$ \\
$const_{FPMB}$ & $1.035\pm0.006$  & $1.035\pm0.006$  & $1.035\pm0.006$  & $1.035\pm0.006$\\
$const_{XRT}$ & $0.78\pm0.03$      & $0.84\pm0.04$  &  $0.74\pm0.03$ & $0.86\pm0.04$  \\
Reduced ${\chi}^2$~(dof) & 0.96~(1178) &  0.96~(1177) & 0.96~(1176)  & 0.95~(1175)\\
\hline
&   & Observation~2 &  \\
\hline
$\Gamma$ &$1.71\pm0.03$   &   $1.71\pm0.04$  & $1.79\pm0.01$  & -\\
${\Gamma}_{2}$ & -  &   $ 1.84\pm0.04$     & - &  -        \\
$E_{cut/break/e}$ (keV)  &  $132_{-39}^{+89} $        &  $7.25_{-0.85}^{+2.50}$  &  $24_{-6}^{+23}$  & $49_{-28}^{+65}$\\
$kT\rm_{in}~(keV)$ & - & - & $0.04\pm0.04$ & $0.04\pm0.03$ \\ 
$\tau$  & -  &  - & - & $0.87_{-0.67}^{+1.10}$  \\
$N^{a}$ & $0.0076\pm0.0004$ & $0.0074\pm0.0006$ & $0.0083\pm0.0002$ & $0.0012\pm0.0007$ \\
$const_{FPMB}$ & $1.01\pm0.01$  & $1.01\pm0.01$  & $1.01\pm0.01$  & $1.01\pm0.01$\\
$const_{XRT}$ & $0.86\pm0.05$      & $0.87\pm0.05$  &  $0.81\pm0.04$ & $0.85\pm0.04$  \\
Reduced ${\chi}^2$~(dof) & 1.0~(594) &  1.0~(593) & 1.0~(594)  & 1.0~(592)\\
\hline
\end{tabular}
\\  
{{\bf{Note}}:  
{a $\rightarrow$ Normalization~($N_{}$)
     is in units of $\rm{photons~cm}^{-2}~\rm{s}^{-1}~\rm{keV}^{-1}$ at 1~keV.} \\
 } 
\end{table*}

\begin{figure}
\includegraphics[height=\columnwidth]{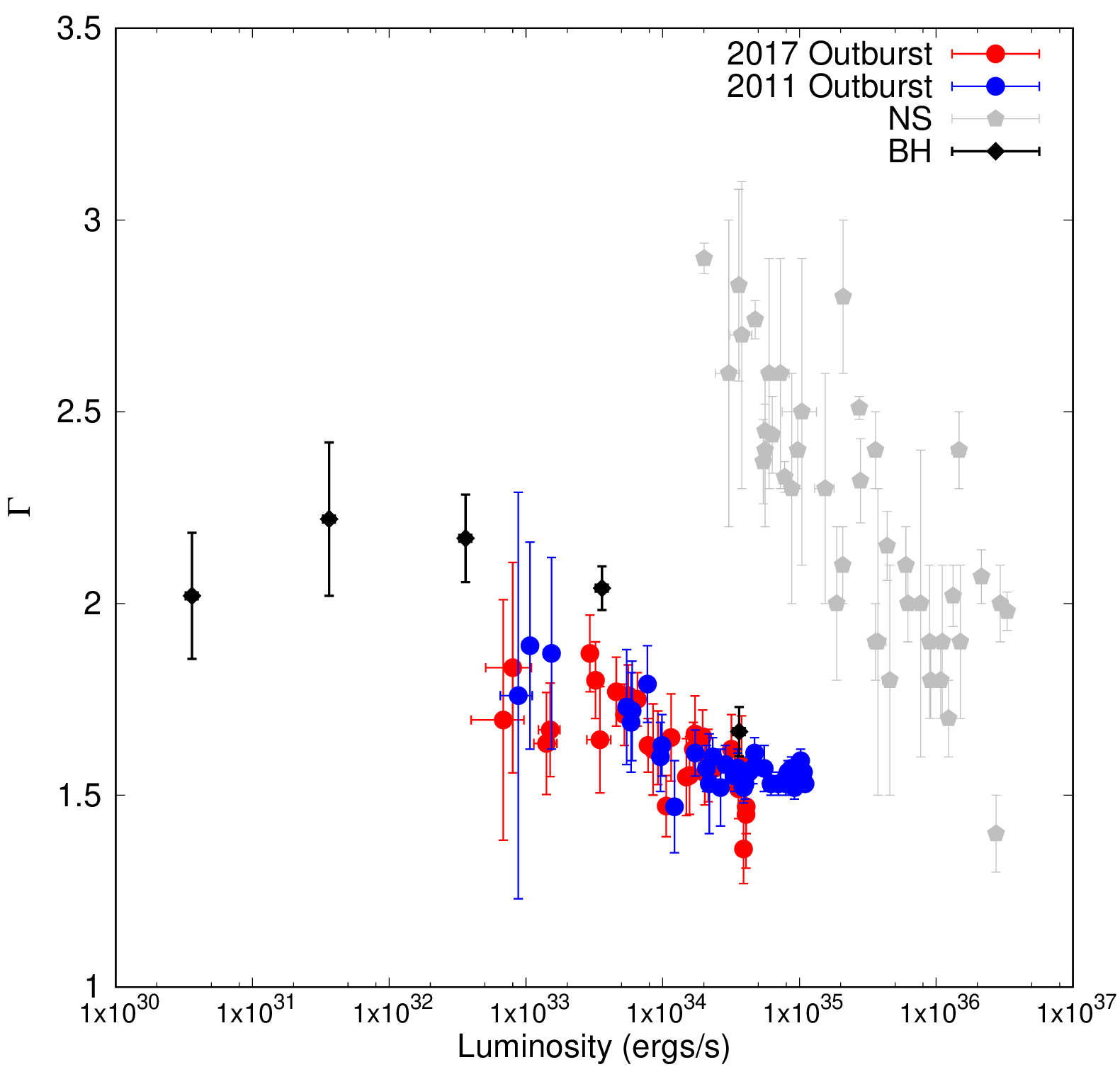}
\caption{The photon index versus the 0.5-10~keV X-ray luminosity for Swift J1357.2-0933, including the data presented by \citet{Wijnands15} for neutron star and black hole binaries. We show measurements made during both 2011 and 2017 outbursts of Swift J1357.2-0933. For the 2011 outburst, we have used the values measured by \citet{Armas13a}.}
\label{Wijnands15}
\end{figure}

\subsection{Spectral Results}

In Figure~\ref{PL-index}, we show the 0.5--10~keV \emph{Swift}-\textsc{XRT} light curve. The maximum of the outburst was observed on $\sim$57864~MJD (OBSID~31918049). The peak value of unabsorbed flux is $\sim$$1.5 \times 10^{-10}~\rm{ergs~cm}^{-2}~\rm{s}^{-1}$ which corresponds to a peak X-ray luminosity of about $4 \times 10^{34}~\rm{ergs}~\rm{s}^{-1}$ assuming a source distance of 1.5~kpc \citep{Rau11}.~We used this value of the source distance in order to compare our results with that obtained during its previous outburst \citep{Armas13a}.~After this, the X-ray luminosity decreased monotonically.~{\bf About}~130 days after the peak our monitoring stopped but the source was still detected during our last observation, albeit at a low luminosities. \\

The middle panel of Figure~\ref{PL-index} shows the power law index~($\Gamma$) evolution over the course of the outburst. We observe that $\Gamma$ increases from a value of $\sim$1.5 to a value of $\sim$1.9. This suggests a softening of the X-ray spectrum. \\

\citet{Wijnands15} searched the literature for reports on the spectral properties of NS and BH LMXBs studied using an absorbed \textit{power-law} model.
They compared the spectra of NH and BH transients 
when they have accretion luminosities between $10^{34}-10^{36}$~$\rm{ergs~s^{-1}}$.
The authors found that neutron star LMXBs are significantly softer than black holes below an
X-ray luminosity~(0.5-10~keV) of $\sim$~$10^{35}$ $\mathrm{ergs~s^{-1}}$ \citep[Figure~1
  of][]{Wijnands15}. In Figure~\ref{Wijnands15}, we plot two outbursts in J1357 together with the data from Figure~1 of \citet{Wijnands15}.  During both the outbursts of J1357, the power-law index showed 
a similar behavior and the data points clearly follow the general trend of the BH sample. \\ 
 To perform simultaneous \emph{NuSTAR}/\emph{Swift}-\textsc{XRT} spectral fitting
 we have used 0.3-79~keV band for the first simultaneous observation
 while for the second observation the 0.7-79~keV band was used.
 A different energy in the lower band was used for the spectral fitting in these observations because the first  contemporaneous \textsc{XRT} observation
 was done using the PC mode while for the second contemporaneous \textsc{XRT} observation, the data {\bf were} collected in the WT mode.
 We first employed an absorbed power-law model for the spectral fitting~(Figure~\ref{spec}).~The residuals obtained after spectral fitting neither showed the presence of any absorption/emission features nor the requirement of any other high-energy component~(e.g.,a Compton hump).~The value of power-law index observed during the first and second observation are $1.66\pm0.01$ and $1.78\pm0.02$, respectively, indicating that the source was in its hard spectral state \citep{Lewin06}.~The absorbed flux measured in the 0.3--10~keV band is $\sim$$3.5 \times 10^{-10}~\rm{ergs~cm}^{-2}~\rm{s}^{-1}$ for the first observation, while it decreased by a factor of $\sim$~2.6 during the second observation.~The best-fit parameters are given in Table~\ref{best-fit}.~We note that the flux reported by \citet{Stiele18} during the first observation is, for unclear reasons, $\sim$~6 times higher compared to that measured in this work. Our measurements are consistent with that reported by \citet{Mondal19}.  \\

We next tried a two-component model (\textsc{diskbb + power law}) and obtained a very low value of the disc temperature~($\sim$~0.03~keV).~However, we find that the additional disc component did not significantly improve the fit. There was no change in the value of 
 chi-squared~(${\chi}^2$) for 2 degrees of freedom~(d.o.f) less.  
\citet{Armas13a} also found that none of the \textsc{XRT} observations required a disc component to fit spectra during the 2011 outburst of J1357. \\

We have also tried the same phenomenological models as suggested by \citet{Stiele18} to compare our spectral fits with those reported earlier by these authors.~Table~\ref{compt}  gives the fit parameters.~Using a cut-off power law to the \emph{Swift} and \emph{NuSTAR} spectra, we found that the obtained cut-off energy lies outside the energy range covered by the data and a
cut-off power law is statistically not required.~The value of reduced ${\chi}^2$ did not change.~Similar results were reported by \citet{Stiele18} but our parameter values are not the same.  
We also found that using a broken power law or an absorbed disc blackbody plus thermal Comptonization model did not further improve the spectral fits.~Our spectral fits suggests that an additional accretion disc component is not required. This is in contrast with that reported earlier by \citet{Stiele18}, might be due to the higher value of flux obtained by these authors. \\

The presence of a neutral iron $K_{\alpha}$ line at 6.4~keV is statistically
not required, however, we obtained an upper limit on the equivalent width~(EW) of $\lesssim$~4~$\mathrm{eV}$  at a 1~$\sigma$ confidence level. We have used an absorbed powerlaw plus Gaussian~(\texttt{tbabs*(powerlaw+gaussian)}) for the spectral fitting and assumed the line width of 10~eV.  \\

To place an upper limit on the reflection fraction in this source we have used a slab model `\textsc{pexrav}'.
\textsc{pexrav} describes X-ray reflection off an infinitely
thick flat slab, from a central illuminating source \citep[][]{Magdziarz95}.~It includes a reflection scaling factor~(R) which gives an estimate of the fraction of X-rays that get reflected off the disc, taking into account other parameters such as the power law 
photon index, cut-off energy, abundance of elements heavier than helium, iron abundance and inclination angle of the slab. We fixed the values of the neutral hydrogen column density to $0.012 {\times}10^{22} \rm{cm^{-2}}$ and power-law index to the best-fit values and assigned a very high value to the cut-off energy~(500~keV)~based on the high value of cut-off energy obtained using an absorbed cut-off power law model.
The inclination angle was fixed to the value 0.63~(cosine of $70^{\circ}$).~This allowed us to attain an upper limit on the reflection parameter~(R) $\lesssim 0.04$ at a 1~$\sigma$ confidence level.~This low value of reflection fraction indicates almost negligible reflection off the disc. \\

Following \citet{Stiele18}, we tried to use a sophisticated relativistic model `\textsc{relxill}' \citep{Dauser14, Garcia14}.~For the first observation, we used an absorbed disc blackbody plus relxill model, however, for the second observation a disc component was not required. 
As in \citet{Stiele18}, we fixed the values of emissivity index to 3, the cut-off energy at 300 keV and outer disk radius ($R_{out}$) at 400~$R_g$.  We also found that it was difficult to constrain ionisation parameter~($log{\zeta}$) and the iron abundance~($A_{Fe}$), if kept free.~Therefore, we fixed the value of 
$log{\zeta}$ =  1 and $A_{Fe}$=0.8 based on the values obtained by \citet{Stiele18}. 
The authors suggested that assuming a low inclination of 
$30^{\circ}$ and a spin value of $\geq$ 0.9 leads to a value of inner disc radius
which is consistent with the expected value at low luminosities~($0.15~{\%}$ of the Eddington Luminosity~($L_{Edd}$)).~Therefore, 
we tried an inclination of $30^{\circ}$ for different values of the spin parameter~(a) 0, 0.8 and 0.9. However, no change in the fit parameters was found on using a different value of spin parameter (see Table~\ref{relxill}). 
Using the above-mentioned values, we observed that using a relxill model did not improve the fit and we
were not able to constrain the inner disc radius and obtained a very low value of the reflection fraction.~\citet{Stiele18} also reported that an inclination of $70^{\circ}$ results into the truncation radius which is quite close to the BH and this contradicts the observed scenario which suggests that the disc is truncated far away from the BH at low luminosities (see their discussion for details).

\begin{table*}
\caption{Spectral fit parameters using an absorbed disc blackbody plus relxill model. For each spectral fit we have used $N\rm{_{H}}$ = $0.012{\times}10^{22}~\rm{cm^{-2}}$. }
\label{relxill}
\begin{tabular}{ c c c c c c c c c}
\hline
\hline
Parameters & a=0 & a=0.8 &  a=0.9 \\
\hline
&   &  Observation~1 (Inclination: $30^{\circ}$ ) &  \\
\hline
$kT\rm_{in}~(keV)$                  &    $0.029_{-0.029}^{+0.002} $                   &    $0.029_{-0.029}^{+0.002} $                        & $0.029_{-0.029}^{+0.002} $    \\
$\Gamma$                                     & $1.63\pm0.01$  &   $1.63\pm0.01$     & $1.63\pm0.01$   \\
$R_{(in)}$$ (R_{g})$  &     $80\pm80$                    &            $80\pm80$                     &              $80\pm80$   \\              
$R_{refl}$                         &    $<$0.03                     &         $<$0.02                         &                 $<$0.02                   \\
$N^{a}$                        & $0.0224\pm0.0004$ & $0.020\pm0.001$ & $0.0239\pm0.0002$  \\
$const_{FPMB}$ & $1.035\pm0.006$  & $1.035\pm0.006$  & $1.035\pm0.006$  \\
$const_{XRT}$ & $0.78\pm0.03$      & $0.84\pm0.04$  &  $0.74\pm0.03$ &  \\
Reduced ${\chi}^2$~(dof) & 0.96~(1175) &  0.96~(1175) & 0.96~(1175)  \\
\hline
&   &Observation~2 (Inclination: $30^{\circ}$ ) &  \\
\hline
$kT\rm_{in}~(keV)$                  &    -                  &    -                        &   -    \\
$\Gamma$                                     & $1.75\pm0.01$  &   $1.75\pm0.01$     & $1.75\pm0.01$   \\
$R_{(in)} (R_g)$                         & $20\pm20$         & $20\pm20$       & $20\pm20$      \\
$R_{refl}$                                      &    $0.01_{-0.01}^{+0.04}$    &   $0.01_{-0.01}^{+0.04}$   &  $0.01_{-0.01}^{+0.04}$  \\
$N^{a}$                        & $0.000204\pm0.000005$ & $0.000204\pm0.000005$ & $0.000204\pm0.000005$ \\
$const_{FPMB}$ & $1.01\pm0.01$  & $1.01\pm0.01$  & $1.01\pm0.01$  \\
$const_{XRT}$ & $0.83\pm0.04$      & $0.83\pm0.04$  &  $0.83\pm0.04$ &  \\
Reduced ${\chi}^2$~(dof) & 1.01~(593) &  1.01~(593) & 1.01~(593)  \\
\hline
\end{tabular}
\\  
{{\bf{Note}}:  
{a $\rightarrow$ Normalization~($N_{}$)
     is in units of $\rm{photons~cm}^{-2}~\rm{s}^{-1}~\rm{keV}^{-1}$ at 1~keV.} \\
 } 
\end{table*}



\subsection{Ultraviolet/Optical and X-ray correlation:}
The simultaneous \textsc{XRT} and \textsc{UVOT} observations allowed us
to study the correlation between the X-ray and the UV/optical emission along the outburst.~Figure~\ref{PL-index}~shows the X-ray light curve
and UV/optical magnitudes in the Vega system. Over the course of the outburst the brightness in all bands decreased with the decline in X-rays.
Following \citet{Armas13a} we fitted these correlations with a power law 
to calculate the correlation slopes~($\beta$, $F_{\rm{UV/optical}}~{\propto}~ F_{\rm{X}}^{\rm{\beta}}$.~The results are given in Table~\ref{beta}. \\


The disc-instability model (including irradiation) predicts the light-curve for the outburst of an irradiated accretion disc to display a characteristic shaped decay profile after the outburst peak \citep{dubus1999,dubus2001}. This profile, which has been observed in a number of BH-LMXB outburst light-curves, is clearly seen in the outburst light-curves of J1357 (e.g., \citealt{Bailey18}).~Therefore,
we compared the correlation between the UV/optical and X-ray fluxes against correlations observed for three emission processes: X-ray reprocessing
in the disc, the viscously heated disc and jet emission. \\

\begin{figure*}
\centering
\begin{minipage}{0.45\textwidth}
\includegraphics[height=3.in,width=8.cm,angle=0,keepaspectratio]{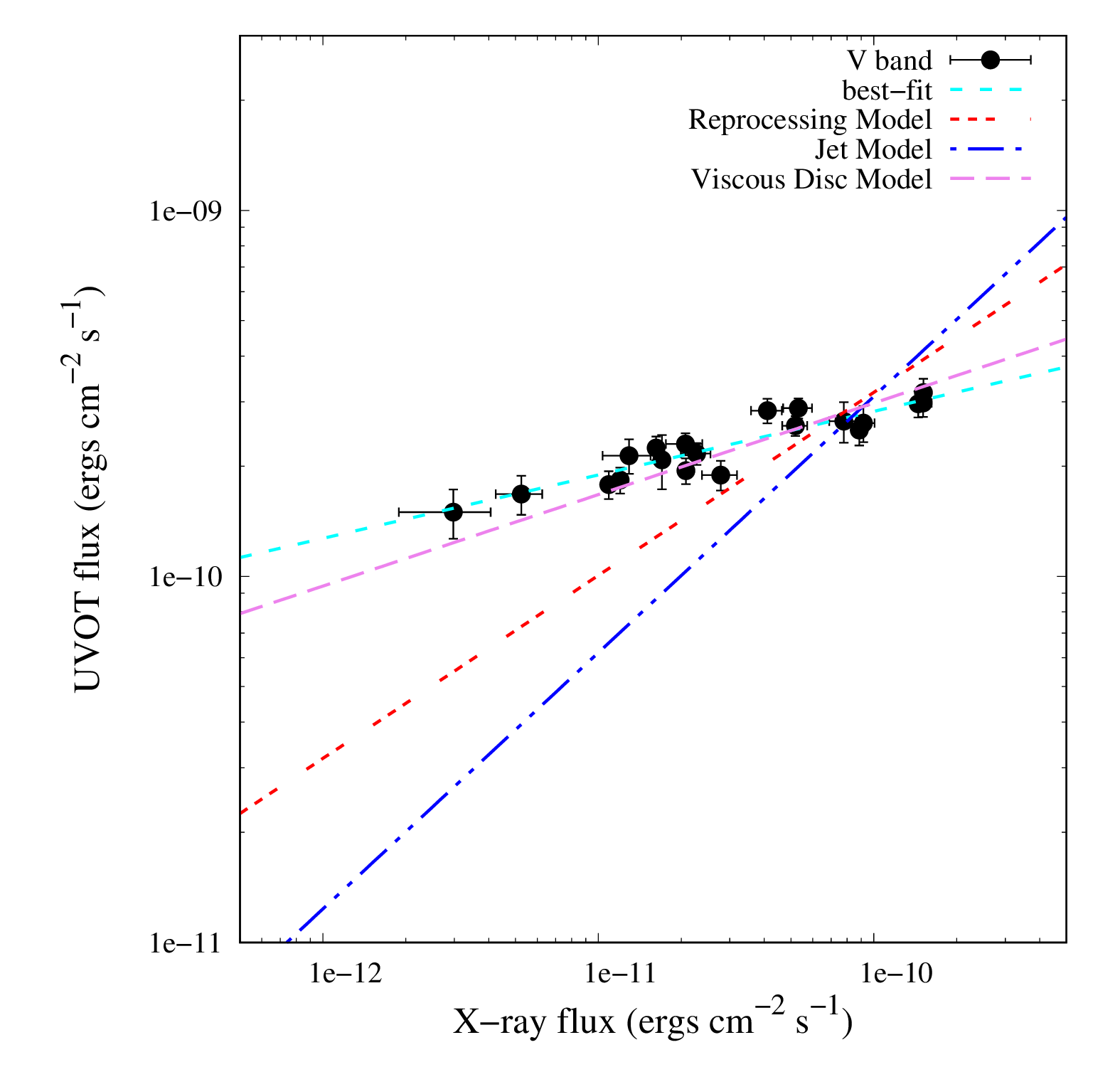}
\includegraphics[height=3.in,width=8.cm,angle=0,keepaspectratio]{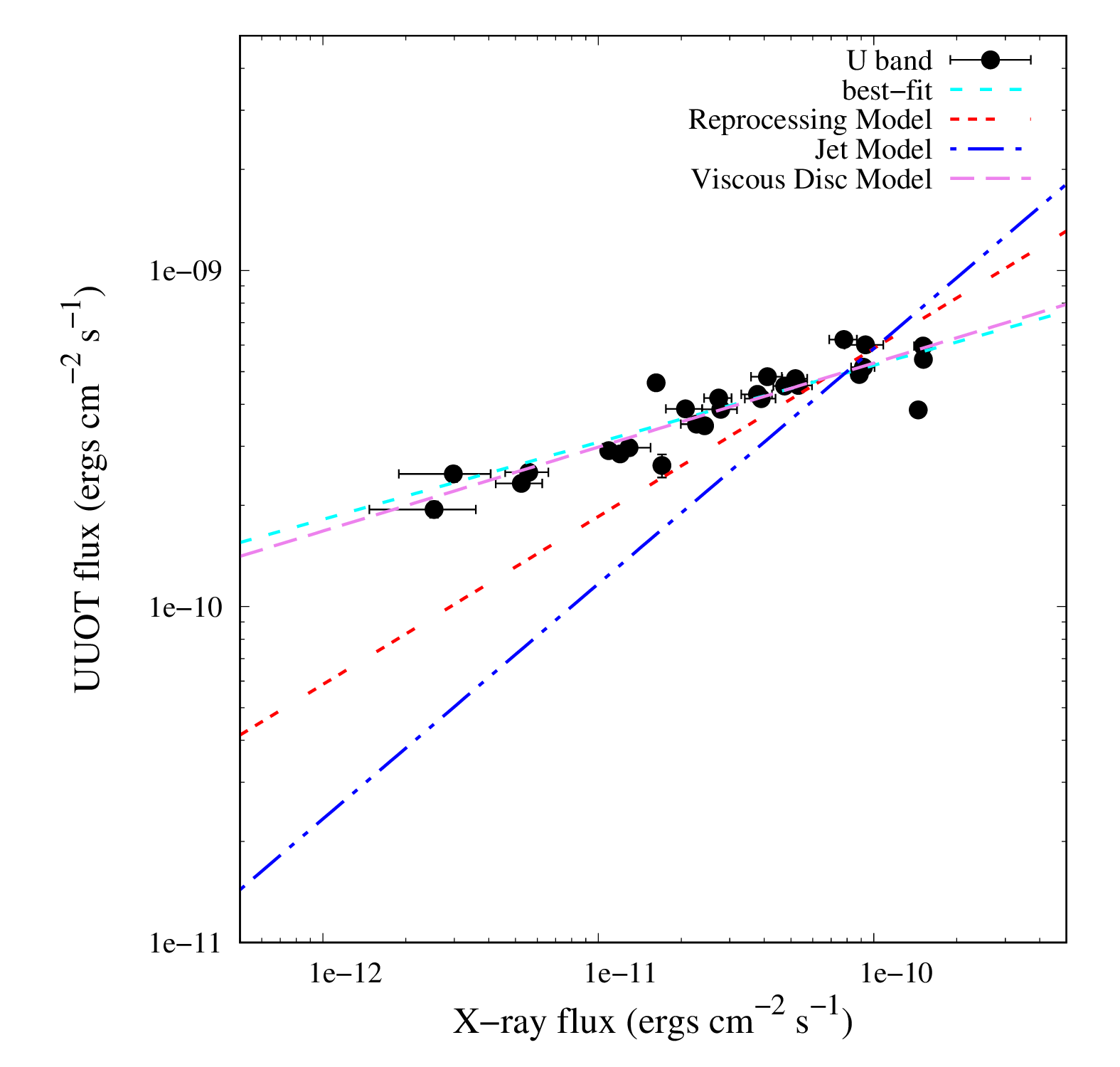}
\includegraphics[height=3.in,width=8.cm,angle=0,keepaspectratio]{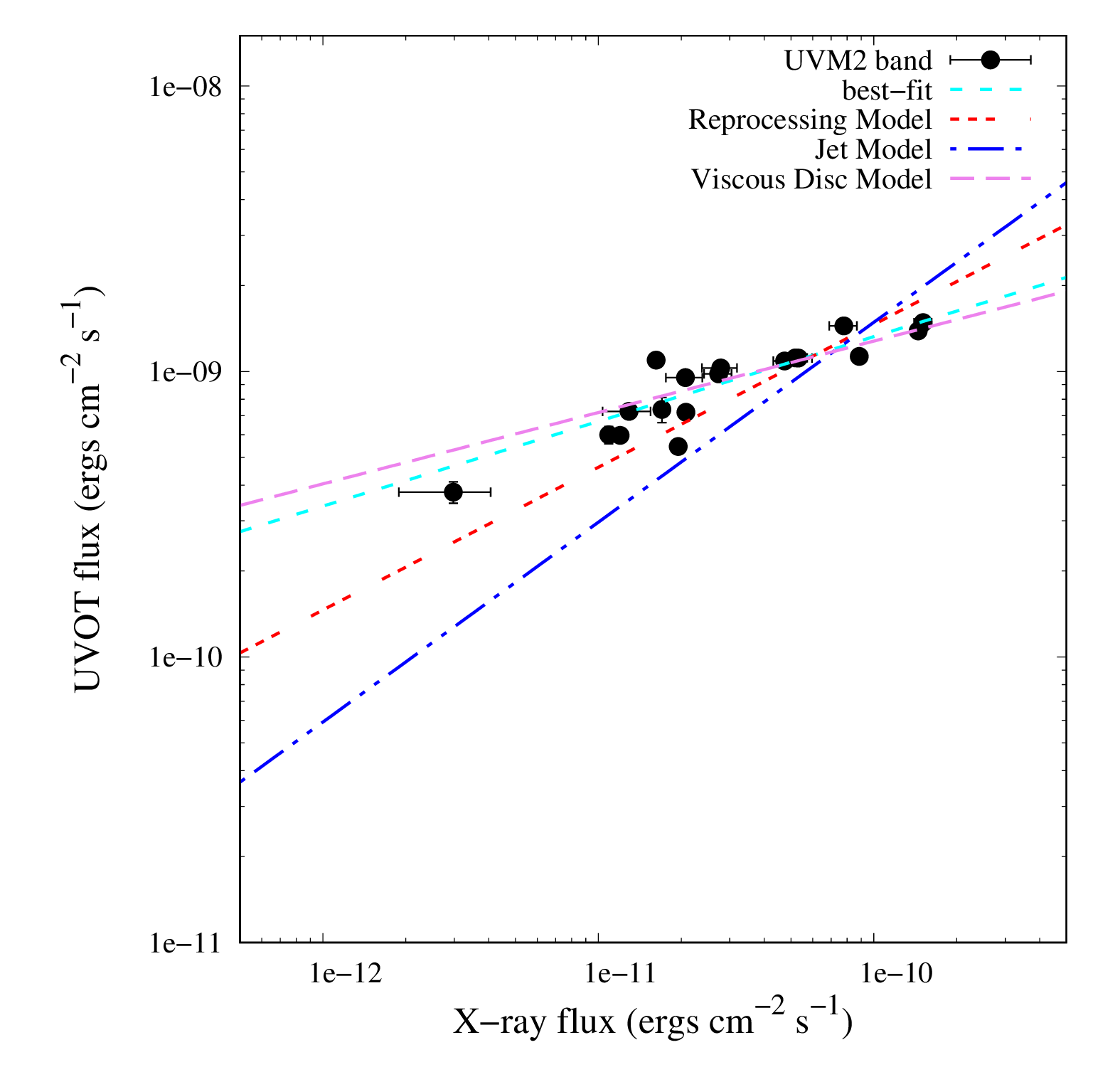}
\end{minipage}
\hspace{0.02\linewidth}
\begin{minipage}{0.45\textwidth}
\includegraphics[height=3.in,width=8.cm,angle=0,keepaspectratio]{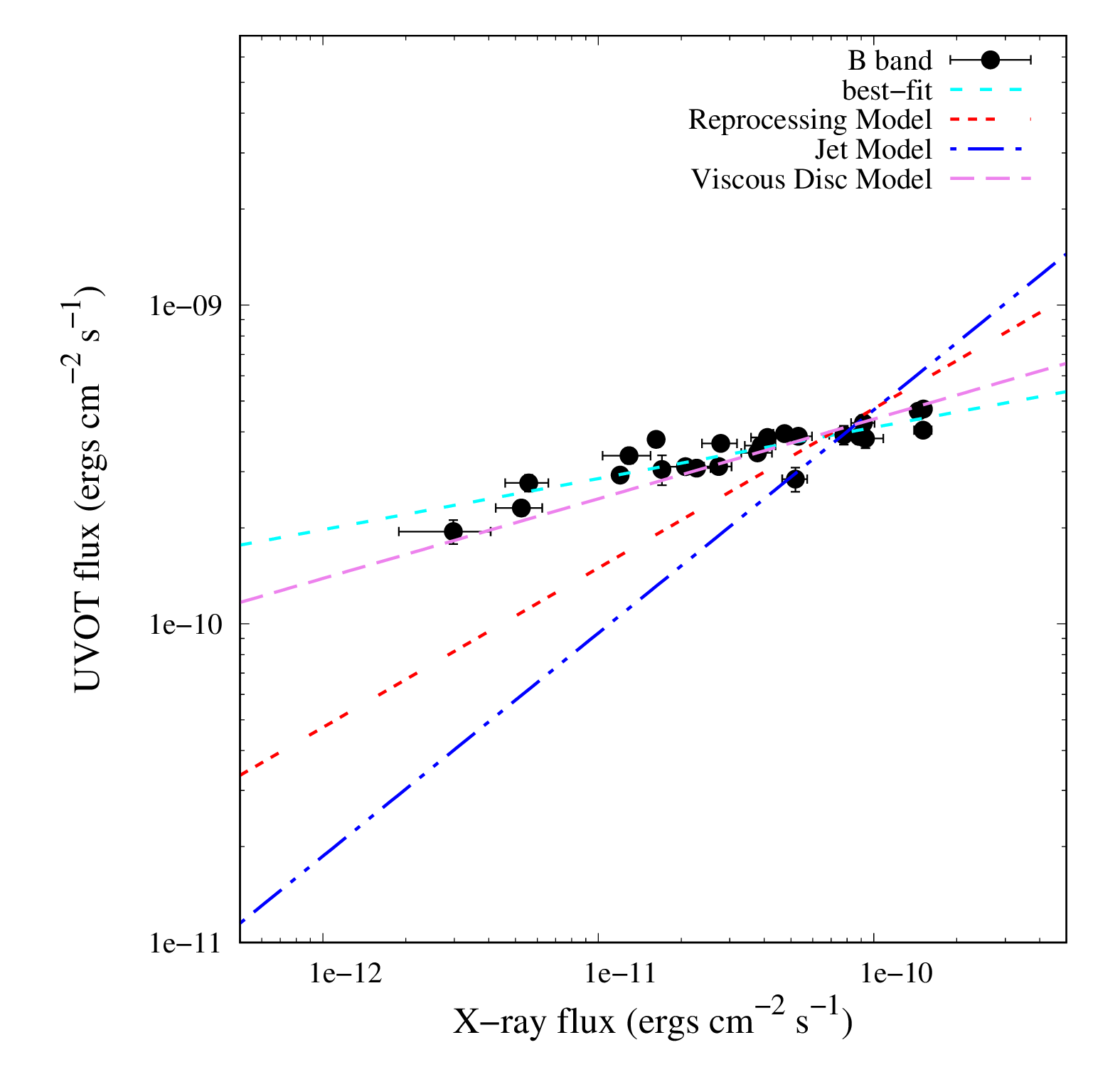}
\includegraphics[height=3.in,width=8.cm,angle=0,keepaspectratio]{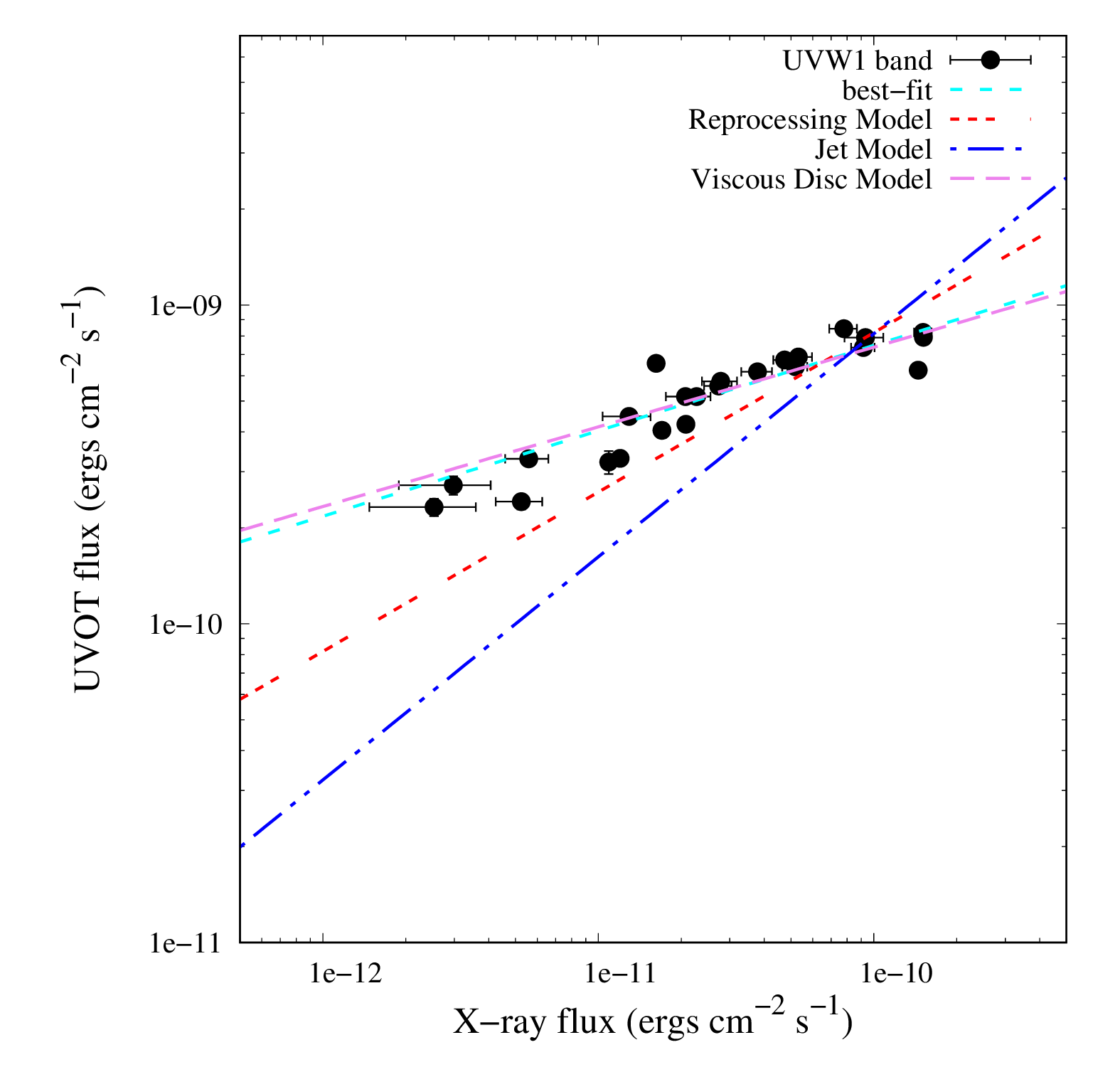}
\includegraphics[height=3.in,width=8.cm,angle=0,keepaspectratio]{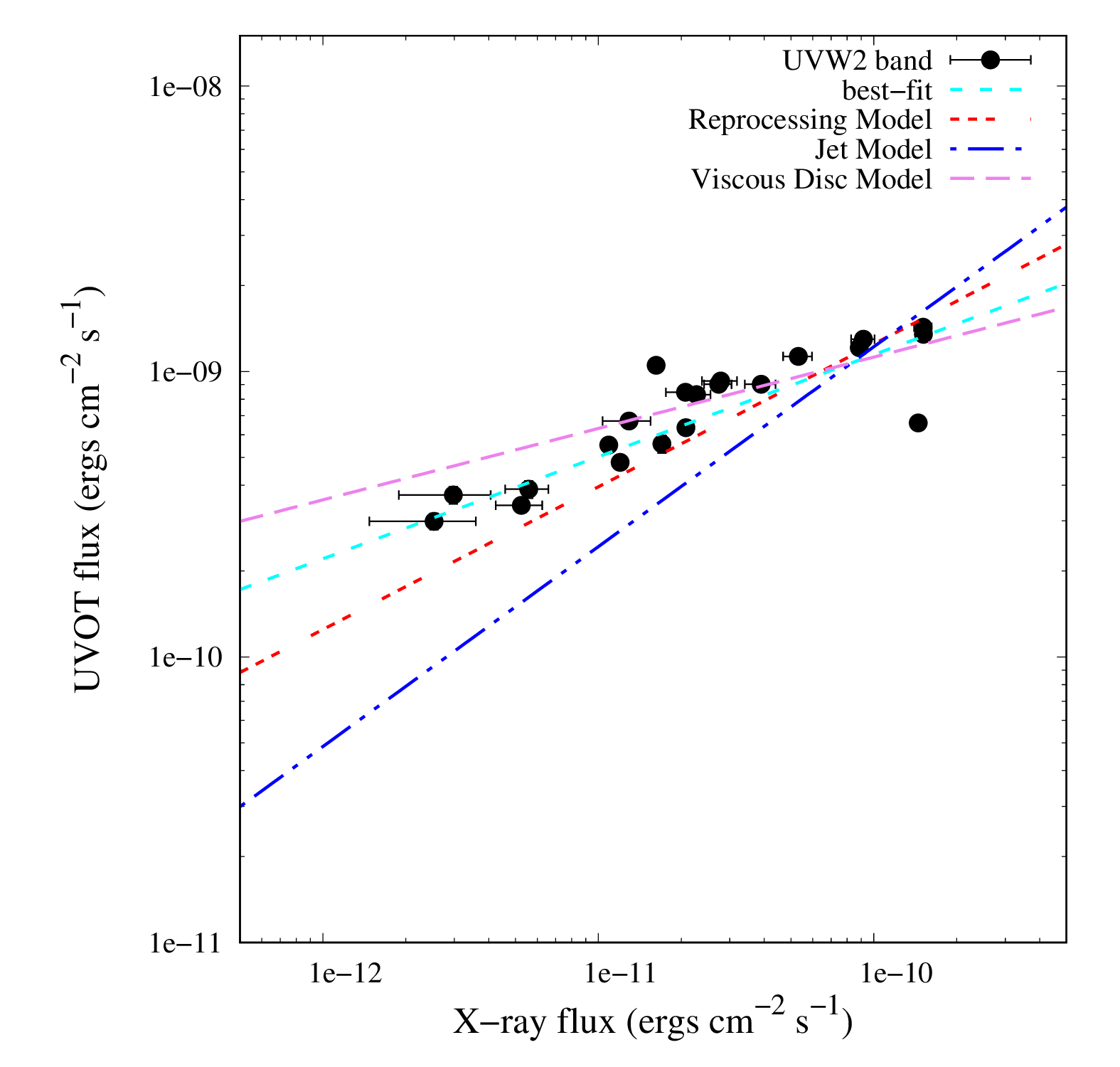}
\end{minipage}
\caption{Correlation slopes (${\beta}$) between the UVOT bands and the unabsorbed X-ray flux in the 2--10 keV energy band, with models for X-ray reprocessing, viscously heated disc and jet components.}
\label{Beta-models}
\vspace{-0.8913pt}
\end{figure*}

\begin{figure}
 \includegraphics[height=\columnwidth,angle=270]{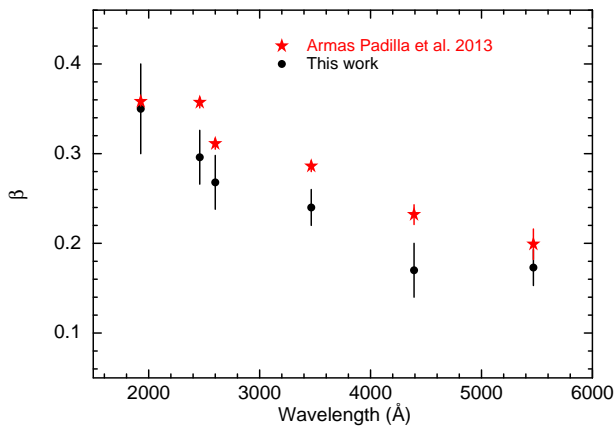}
 \caption{Best-fitted correlation slopes (${\beta}$) between the UVOT bands and the X-ray flux in the 2--10 keV energy band.}
 \label{correlation}
 \end{figure}

For X-ray reprocessing, we adopt the theoretical model between the optical and X-ray luminosities given by
\citet{Van94}. This model predicts that the optical luminosity of an X-ray reprocessing accretion disc varies as $L_{opt}$~${\propto}$~${L_{X}}^{0.5}a$, where $a$ is the orbital separation of the system given by $3.5\times{10^{10}}(M_{BH})^{1/3}(1+q)^{1/3}(P_{hr})^{2/3}$ \citep{Frank02}. For J1357, we have adopted the values of mass of the BH~($M_{BH}$),~the mass ratio of the companion star to the compact object,~$q=M_C/M_{BH}$ and the orbital period~($P_{hr}$) from \citet{Casares16}.~For the viscously heated disc and jet emission we have used the following relations:~
$L_{opt}\propto~L_{X}^{0.25}$ and $L_{opt}\propto~L_{X}^{0.7}$, respectively \citep[see][]{Russell06}.~Figure~\ref{Beta-models} shows that for all the \textsc{UVOT} bands
our best-fitted correlation slope lies closer to the model for the optical emission arising from a viscously heated disc around a BH. The other two models, namely the X-ray reprocessing and jet component do not fit these data well.~However, for the UV emission it seems that there is not a single process going on in wavelength, very similar to
what \citet{Armas13a} found for the 2011 outburst of the source (their points are plotted as well).
In the Figure~\ref{correlation}, we plotted the correlation slopes obtained for all the \textsc{UVOT} bands which shows an increase in $\beta$ with decrease 
in the wavelength. \\

 We also tried the combined spectral fitting using the optical/UV and X-ray data, however, we find that it was not possible to 
constrain fit parameters using the model `\textsc{diskir}'. Therefore, we do not further comment on this. The broadband spectral
fits have been discussed in detail in Paice et al.~(in prep). 
 \begin{table*}
\caption{Correlation slope between UV/optical and X-ray fluxes during 2017 outburst of J1357. }
      \label{beta}
 \begin{tabular}{cc |ccccccc}
 \hline
       &              &          &          \\ [0.5ex]       
  UVOT Band& Wavelength & $\beta$~(fit values)  \\ [0.5ex] 
\hline
v & 5402 & $0.17\pm0.02$  \\ [0.5ex] 
b & 4329 & $0.17\pm0.03$\\ [0.5ex] 
u & 3501 & $0.24\pm0.02$   \\ [0.5ex]
uvw1 & 2634   & $0.27\pm0.03$  \\ [0.5ex]
uvm2 & 2231 & $0.30\pm0.03$  \\ [0.5ex]
uvw2 & 2030 &  $0.35\pm0.05$ \\ [0.5ex]
\hline
\\
\end{tabular}
\\
{{\bf{Note}}:  
 Errors quoted are {\bf at} 1-sigma confidence. } 
 \end{table*}   


%
%
%


\section{Discussion}
In this work, we report a multi-wavelength analysis of J1357 during its second recorded outburst, obtained using the observations made with the \emph{Swift} and \emph{NuSTAR} observatories.
\subsection{Timing Behavior}
The timing studies performed with the light curves obtained with the \emph{NuSTAR} data did not show any dips or eclipse.
This behavior of the X-ray light curves is consistent with that observed during its previous outburst in 2011 \citep{Armas13b}. ~The power density spectra created 
using data of both the \emph{NuSTAR} observations (which are separated by more than 1 month and 10 days) showed the presence of some excess power in the milli-hertz frequency range. 
We observed that the Q factor of the broad feature at $\sim$ 3~mHz and $\sim$ 0.019~mHz is $\sim$~0.6 and $\sim$~1.0 respectively.~Thus, it seems that the features observed in the PDS are not QPOs \citep{Vander2000}.
 \citet{Armas13b} reported the presence of a QPO at $\sim$~6~$\mathrm{mHz}$ having a Q value of about 3 or larger
 in one of the \emph{RXTE} observations made close to the beginning of the outburst.~However, the \emph{XMM-Newton} observation made 3.5 days after the first \emph{RXTE} observation did not show 
 this QPO. 
 Thus, the absence of a QPO in the \emph{NuSTAR} observations
 made a week after the peak of the outburst is consistent
 with the previous reports by \citet{Armas13b}. \\
 

Based on the discovery of the optical dips, \citet{Corral13} suggested the presence of obscuring torus in the inner accretion disc which moves outwards when the X-ray luminosity decreases.~However, based on the X-ray study of J1357, \citet{Armas13b} question this interpretation and suggested that something else might cause the optical dips. \\

It is observed that high inclination X-ray sources show deep, irregular shaped dips in the X-ray light curves because the accretion disc structure in the outer disc extends vertically above the plane of the binary and periodically blocks the line of sight to the central compact object.~This structure is a sort of bulge at the disc edge, associated with the impact point of the accretion stream from the secondary star with the outer accretion disc.~Since J1357 has been suggested to be a high inclination binary system, we searched the literature for any possible similarities of this source with any other X-ray dipping sources.~J1357, however, behaves quite differently from X-ray dipping sources.~One of the X-ray dipping sources,  4U~1254--690,~shows cessation of X-ray dipping activity~\citep{Smale99}.
4U~1254--690 is a low mass X-ray binary having an inclination angle of $65^{\circ}-73^{\circ}$ \citep{Motch87}.~This source usually displays dips during which the observed reduction in flux is 20$\%$-95$\%$ \citep{Courvoisier86}.~However, during observations made with \emph{RXTE} in 1997~(May) did not show the presence of dips. The non-existence of dips in the X-ray light curves of 4U~1254--690 were explained due to shrinkage of the angular size of the bulge height \citep[from 17-25~$\%$ to less than 10$\%$;][]{Smale99}.
 Therefore, it might be possible that non-existence of dips in the X-ray light curves of J1357 is due to a low angular size of the bulge height~(less that 10$\%$) as predicted for 4U~1254--690.~In that case, J1357 spends most of its time with this configuration and hence we observe no dips in the X-ray light curves.~\citet{Galloway16} reported on sporadic dipping in Aql~X--1 and suggested systems with high inclination can exist despite not showing persistent dipping activity.

\subsection{Spectral Behavior} 
We found that a simple absorbed power law was adequate to fit the spectra 
obtained with the observations made with \emph{Swift}-\textsc{XRT} and \emph{NuSTAR} during the 2017 outburst of J1357. The observed power-law index showed that the source was in its hard spectral state during all the observations~(Table~\ref{Swift},\ref{best-fit}).
The evolution of power-law index with the X-ray luminosity follow a similar trend during the 2011 and 2017 outbursts of J1357.  \\




The \emph{Swift}-\textsc{XRT} observations performed during the 2017 outburst of J1357 did not reveal the presence of any soft excess such as seen during an \emph{XMM-Newton} observation
obtained during the 2011 outburst of J1357 \citep{Armas13b}. This might again be due to the limitations on the data \citep[see also][]{Armas13b}.
It is believed that during the low/hard spectral state of black hole binaries~(BHBs), the disc is either not detected \citep[see e.g.,][]{Belloni99}, or it appears much cooler and larger than it does in the soft state \citep[see][]{Wilms99,McClintock01,Reynolds13}.~Therefore, it is not possible to observe such a cool disc with the instruments like \emph{NuSTAR} that have a low energy cut-off of approximately 3~$\mathrm{keV}$.
\citet{Reis10} studied eight BHBs during their low hard state using the data from \emph{XMM-Newton}, \emph{Suzaku} and \emph{Chandra} observatories.~They found the presence of thermal emission in all of these sources.~They found the disc temperature as low as $\sim$ 0.2~$\mathrm{keV}$. 
Thus, the non-detection of the soft component in J1357 suggests that it was in its hard spectral state. \\

Signatures of reflection in the form of an iron line and Compton hump
are often observed in the X-ray spectra of BHBs during their low/hard state \citep[e.g.,][for reviews]{Miller06,Reynolds10}.
The reflection components are however
most apparent when the accretion disc is observed nearly face-on \citep[see e.g.,][]{DiSalvo01,Tomsick09, Furst15, Furst16}.
J1357 is proposed to have a torus-like structure in the inner region of accretion disc similar to the torus observed in
several Compton-Thick~(CT) active galactic nuclei~(AGN).~Observations suggest that the X-ray spectrum of CT AGNs is dominated by the
cold Compton reflection component which arises from the Compton scattering of the inner ``wall'' of the neutral obscuring torus.
This emission is characterised by a hard X-ray spectral slope with a peak around 30~keV as well as high EW of fluorescent emission line like the 6.4~keV line \citep[see e.g][]{Arevalo14,Gandhi14,Bauer15}).~From the broad band spectral study performed with data of \emph{Swift}-\textsc{XRT} and \emph{NuSTAR} we did not find the presence of any reflection features~(iron line or Compton hump) that would support the presence of clumpy torus in the disc.~An upper limit of $\lesssim$0.04 on the reflection fraction was obtained using the slab model~\textsc{`pexrav'}.~We also note that using a relativisitic `\textsc{relxill}' model did not improve the spectral fits and the measured values of the reflection fraction are very low and unconstrained.  \\

We did not observe an iron line in the X-ray spectra of J1357 at luminosities 
of about $\sim$~$(2-9){\times}10^{34}~\mathrm{ergs~s^{-1}}$.~However, 
there have been several reports  of  broad  iron  lines  in  the  brighter  part  of  the  low/hard  state (greater than $10^{36}~\mathrm{ergs~s^{-1}}$) during an outburst~for example, GX 339--4 \citep{Miller06,Reis10}, GRS~1739--278 \citep{Miller15} and as well as for 
many other systems \citep[see][for details]{Reynolds10}.~The typical range of iron line EW is 50-300~eV \citep{Gilfanov10}.~A detailed study of GX~339--4 in the low/hard state at a luminosity of about $10^{35}~\mathrm{ergs~s^{-1}}$ showed the presence of
a narrow iron line in the X-ray spectrum \citep{Tomsick09}.
The EW of the observed iron line is $\sim$~77~$\mathrm{eV}$.~Based on their study, the authors suggested the truncation of the accretion disc for stellar mass black holes in the hard state, at low luminosities~($10^{34}-10^{36}$~$\mathrm{ergs/s}$).
They also found the drop in the iron line EW with the increase in the inner accretion disc radius~($R_{in}$).
Another detailed spectral study on GRS~1739--278 during its low hard state showed the presence of reflection features in the spectrum at the luminosity of about $\sim$~$2{\times}10^{34}~\mathrm{ergs~s^{-1}}$ \citep{Furst16}. 
For the advection dominated accretion flow~(ADAF) one requires a large increase in $R_{in}$ with decreasing $L_X$.~Therefore, it might be possible that $R_{in}$ for the case of J1357 is quite large and because of this we do not observe reflection features in the X-ay spectrum. \\






\

\subsection{UV/optical and X-ray Correlation}

During the 2017 outburst of J1357, we found a set of values 
of the correlation coefficient~($\beta$; Table~\ref{beta}).~We observe that the value of the correlation coefficient in the v band is $\sim$ 0.17. This value lies between $0.15{\leq}{\beta}{\leq}0.25$ which is expected for a BH system with a viscously heated accretion disc \citep{Armas13a}.~This is supported by the fact that the best-fitted correlation slopes~($\beta$) can be explained by the model for a viscously heated disc.  \\

We also observe a clear increase in the value of $\beta$ with the decrease in wavelength~(Figure~\ref{correlation}) as also observed by \citet{Armas13a}.~The best-fit values of $\beta$ for each band deviate largely from the predicted values for the reprocessing model and the jet model but it might be possible
that intrinsic thermal emission from the viscously heated outer accretion disc contribute significant light in the optical \citep{Frank02}.~Moreover, it is suggested by \citet{Van94} that
for smaller accretion discs (i.e., smaller $P_{orb}$) we expect the average surface temperature of the disc to be larger (presumably as it is closer to the compact object and irradiation source). 
Therefore, we expect a larger fraction of the reprocessed emission to be in the UV band. Given that we observe the slope of the correlation between UV and X-ray emission get steeper (i.e., $\beta$ increases) as {\bf the} wavelength of the UV band used decreases, this could explain the values of $\beta$ between the V band
and X-ray more consistent with the viscous disc only, rather then the irradiated disc.

\section*{Acknowledgments}
We thank the anonymous referee for several useful suggestions which improved the quality of the paper.
 A.B is grateful to the Royal Society and SERB (Science and Engineering Research Board, India) for financial support through Newton-Bhabha Fund.~A.B. is supported by an INSPIRE Faculty grant (DST/INSPIRE/04/2018/001265) by the Department of Science and Technology, Govt. of India.
 She also gratefully acknowledge Dr. Adam~Hill for help in the installation of \emph{NuSTAR} Timing Analysis Software package. DA acknowledges support from the Royal Society. ND and JVHS are supported by a Vidi grant awarded to ND by the Netherlands Organization for Scientific Research (NWO).~MJM acknowledge support from STFC Ernest Rutherford fellowships.~This work is based on data from the \emph{NuSTAR} and \emph{Swift} mission. We would like to thank all the members of \emph{NuSTAR} and \emph{Swift} team for TOO observations. This research has made use of data and/or software
provided by the High Energy Astrophysics Science Archive Research
Center (HEASARC), which is a service of the Astrophysics Science
Division at NASA/GSFC and the High Energy Astrophysics Division of the
Smithsonian Astrophysical Observatory.~This research has made use of
NASA's Astrophysics Data System.

\bibliographystyle{mnras}
\bibliography{main}

\label{lastpage}
\end{document}